\documentclass[12pt]{article}
\usepackage{amsmath,amssymb,amsfonts,epsf}
\usepackage{multirow}
\def\ci{\mathrm{i}}
\def\ud{\mathrm{d}}
\def\Tr{\mathrm{Tr}}

\newcommand{\expp}{\mathrm{e}}
\renewcommand{\Re}{\operatorname{Re}}

\renewcommand{\v}{\boldsymbol v}

\newcommand{\s}{\boldsymbol s}
\newcommand{\q}{\boldsymbol q}
\newcommand{\A}{\boldsymbol A}
\renewcommand{\u}{\boldsymbol u}

\begin{document}
~\vspace{3.0cm}

\centerline{\LARGE Semiclassical expansion of parametric}
\vspace{0.3cm}
\centerline{\LARGE correlation functions of the quantum time delay}
\vspace{2.0cm}
\centerline{\large 
Jack Kuipers\footnote{E-mail: jack.kuipers@bristol.ac.uk}
and Martin Sieber\footnote{E-mail: m.sieber@bristol.ac.uk}}
\vspace{0.5cm}

\centerline{School of Mathematics, University of Bristol,
Bristol BS8\,1TW, UK}

\vspace{5.0cm}
\centerline{\bf Abstract}
\vspace{0.5cm}
We derive semiclassical periodic orbit expansions for a correlation
function of the Wigner time delay. We consider the Fourier transform
of the two-point correlation function, the form factor $K(\tau,x,y,M)$,
that depends on the number of open channels $M$, a non-symmetry breaking
parameter $x$, and a symmetry breaking parameter $y$. Several terms in
the Taylor expansion about $\tau=0$, which depend on all parameters,
are shown to be identical to those obtained from Random Matrix Theory. 

\vspace{2.5cm}

\noindent PACS numbers: \\
\noindent 03.65.Sq ~ Semiclassical theories and applications. \\
\noindent 05.45.Mt ~ Semiclassical chaos (``quantum chaos'').

\clearpage

\section{Introduction}

Random Matrix Theory (RMT) and semiclassical approximations are
two alternative theoretical approaches for quantum systems
that are chaotic in their classical limit. RMT describes universal,
system-independent, statistical quantum fluctuations by modelling
the Hamiltonian with matrices that have random elements 
\cite{Meh04,BGS84,Haa01}.
Semiclassical methods on the other hand yield approximations
in terms of classical trajectories of individual chaotic systems,
asymptotically valid as $\hbar \rightarrow 0$, and can also describe
non-universal system-specific quantum properties \cite{BB97,Sto99}.
In recent years, however, a main emphasis in the application of
semiclassical methods has been to obtain the universal results
of RMT from the asymptotic properties of long trajectories and
thus provide a theoretical foundation for the applicability
of RMT in chaotic systems.

The first quantities that were treated in this way were spectral
correlation functions of closed systems which have a semiclassical
expansion in terms of a double sum over periodic orbits. For
the spectral form factor $K(\tau)$, the Fourier transform
of the two-point correlation function of the density of states,
the terms in the Taylor expansion about $\tau=0$ were derived
semiclassically in agreement with RMT. The first term in the
Taylor expansion was obtained from the diagonal approximation
for the double sum over periodic orbits which pairs orbits
with themselves or their time reverse \cite{HO84,Ber85}. 
Higher order terms are due to pairs of different, but strongly
correlated periodic orbits. The first off-diagonal term, i.e.
the second term in the Taylor expansion, was calculated
for uniformly hyperbolic systems in \cite{SR01,Sie02}. The third
order for general chaotic systems was obtained in \cite{HMBH04},
and all higher order terms in  \cite{MHBHA04,MHBHA05,Mue05}. 
Similar methods have been applied to derive off-diagonal
terms, for example, for the conductance \cite{RS02,HMBH06}, the
GOE-GUE transition \cite{NS06}, the shot-noise \cite{BHMH06} and 
parametric correlations \cite{NBMSHH06,KS06}.

In this article we consider two-point correlation functions
of the Wigner time delay in open systems, for which there are
very comprehensive results from RMT available \cite{LSSS95,FS96,FSS97}.
We evaluate a semiclassical periodic orbit expansion for
the Fourier transform of the two-point correlation function,
the form factor $K(\tau,x,y,M)$. It depends on a (scaled)
non-symmetry breaking parameter difference $x$, the number of open
channels $M$, and parameter $y$ that depends on a time reversal
symmetry breaking magnetic field.
Earlier work treated $K(\tau,M)$ and $K(\tau,y,M)$
in the diagonal approximation \cite{Eck93,VOL98}. 
We derive several terms in the Taylor expansion of $K(\tau,x,y,M)$
around $\tau=0$ and compare them to RMT. We find complete agreement
for the terms in this expansion that depend on the four parameters
$\tau$, $x$, $y$, and $M$.

In section~2 we introduce the form factor $K(\tau,x,y,M)$,
and in section~3 we state results of RMT
for it. Sections~4 and~5 treat the case without
symmetry breaking parameter $y$ and contain derivations
of diagonal and off-diagonal terms for $K(\tau,x,M)$,
respectively. The general case with additional parameter
$y$ is treated in section~6, and section~7 contains our
conclusions. 

\section{The form factor for the time delay}

The Wigner time delay is a measure of the
typical time that is spent in a scattering process in
comparison to free motion. It is defined in terms of
the scattering matrix $S(E)$ as \cite{Wig55,Smi60}
\begin{equation}
\tau_W(E)=-\frac{\ci \hbar}{M}\Tr\left[S^{\dagger}(E)\frac{\ud}{\ud E}S(E)\right] 
= -\frac{\ci \hbar}{M} \frac{\ud}{\ud E} \ln\det S(E) \; ,
\end{equation}
where $M$ is the number of open scattering channels. We allow the
system to depend on external parameters $X$ and $Y$ and will
include this later in the notation. A comprehensive overview
of the history and the applications of the time delay is given
in the review \cite{CN02}.

One important property of the Wigner time delay is that it can
be related to a density of states \cite{Fri52,CN02}
\begin{equation} \label{reldens}
\tau_W(E) = \frac{2\pi\hbar}{M} d(E) \approx
 \frac{2\pi\hbar}{M} \left( \bar{d}(E)+ d^{\text{osc}}(E) \right)
\end{equation}
The density $d(E)$ here is the difference between the level 
density of the scattering system and a free
or reference system, because the level density itself is infinite.
In the semiclassical regime it is divided into a mean part
$\bar{d}(E)$, and an oscillatory part $d^{\text{osc}}(E)$.
The first term in (\ref{reldens}) represents the average
time delay which can be expressed as $T_H/M$ where
$T_H = 2 \pi \hbar \bar{d}(E)$ is the Heisenberg time.
We will discuss later that the oscillatory part is
semiclassically approximated by the periodic orbits in
the repeller. In the following we will consider
the case of a two-dimensional
open chaotic cavity with attached leads. Then $\bar{d}(E)$
is given by Weyl's law for the closed cavity 
$\bar{d}(E) \sim \Omega/(2 \pi \hbar)^2$, 
where $\Omega$ is the volume of the surface of constant energy $E$
in phase space, and $d^{\text{osc}}(E)$ is given in
terms of the periodic orbits in the open cavity \cite{VOL98}.

Quantum fluctuations of the Wigner time delay are 
characterized by its correlation functions. If one compares
the time delay at different energy and parameter values
one has to scale the energy difference as well as the
parameter difference in order to obtain universal correlation
functions in chaotic systems. Similarly as for closed systems
one can define a new energy parameter $\tilde{E}$ by
\begin{equation} \label{unfolde}
\tilde{E} = \bar{N}(E,X) \; ,
\end{equation}
where $\bar{N}(E,X)$ is the mean part of the integrated level density
$N(E,X) = \int_{-\infty}^E \ud E' \; d(E',X)$. 
In term of this `unfolded' energy variable the level density has
a mean value of one. Small intervals of the two energies
are related by $\Delta \tilde{E} \approx \Delta E / \bar{d}(E)$.
In closed systems, the external parameter $X$ is unfolded by
requiring that the variance of the level velocities $\partial E_n
/\partial X$, of the energy levels $E_n$, is unity in terms of the
new parameters $\tilde{X}$
and $\tilde{E}$. One then obtains $ \Delta \tilde{X} \approx
\Delta X / \sigma$, where $\sigma^2$ is the variance of the level
velocities. For open systems we scale the parameter difference
again by a quantity $\sigma$ whose value we will discuss later
(see equation (\ref{semisigma})). 

In the semiclassical limit one can then define a universal 
correlation function by (see \cite{KS06} for closed systems)
\begin{equation} \label{r2}
R_2(\eta,x,y,M) \sim \frac{\left\langle
\tau_W^\text{osc}\left(E + \frac{\eta}{2 \bar{d}} + \frac{x \rho}{2 \sigma}
,X + \frac{x}{2 \sigma},Y \right) \;
\tau_W^\text{osc}\left(E - \frac{\eta}{2 \bar{d}} - \frac{x \rho}{2 \sigma},
X - \frac{x}{2 \sigma},Y \right) \;
\right\rangle_{E,X}}{\bar{\tau}_W(E,X)^2}
\end{equation}
where $y$ is a function of $Y$ to be specified later. The average
is carried out over an energy interval $\Delta E$, satisfying
$E \ll \Delta E \ll 1/\bar{d}(E)$, and over a parameter interval
$\Delta X$. We note that changes of order one of $x$ and $\eta$
correspond to very small changes on the classical scale. The term
$x \rho /2 \sigma$ accounts for the change of the energy when $X$
is changed while keeping $\tilde{E}$ fixed
\begin{equation}
\rho = \left. \frac{\partial E}{\partial X} \right|_{\tilde{E}} =
- \frac{\partial \bar{N} / \partial X}{\partial \bar{N} / \partial E} \; .
\end{equation}
We can also express the two-point correlation function in terms of 
the level density
\begin{equation} \label{r2tau}
R_2(\eta,x) \sim \frac{\left\langle
d^{\text{osc}}\left(E + \frac{\eta}{2 \bar{d}} 
+ \frac{x \rho}{2 \sigma},X + \frac{x}{2 \sigma} \right) \;
d^{\text{osc}}\left(E - \frac{\eta}{2 \bar{d}} - \frac{x \rho}{2 \sigma},
X - \frac{x}{2 \sigma} \right) \;
\right\rangle_{E,X}}{\bar{d}(E,X)^2}
\end{equation}
 
The parametric form factor for the time delay is obtained by a Fourier
transform of this two-point correlation function
\begin{equation} \label{defform}
K(\tau,x) = \int_{-\infty}^\infty R_2(\eta,x) \, e^{-2 \pi i \eta \tau}
\; d \eta \; .
\end{equation}

\section{Results from Random Matrix Theory} \label{sect_rmt}

In this section we review the RMT calculations for the form factor
for the time delay.  We are interested in the small $\tau$ expansion of the form
factor and we restrict ourselves to the case $\tau<1$ in the following. 
We consider first the results for the case where we do not have a parameter
that breaks time reversal symmetry, and state the results for the 
Gaussian Orthogonal Ensemble (GOE) and the Gaussian Unitary Ensemble (GUE)
that are relevant for systems with and without time reversal symmetry,
respectively. The semiclassical result will be derived for perfect transmission
$T=1$ in all scattering channels, and we state the RMT results only for this case.

The parametric two-point correlation function $R_2(\eta,x,M)$ for the GUE case
was derived in \cite{FS96} and is given by
\begin{equation}
R_2^{\text{GUE}}(\eta,x,M) = \frac{1}{2} \int_{-1}^1 \ud \lambda \int_1^\infty \ud \lambda_1 \; 
\cos(\pi\eta(\lambda_1-\lambda)) {\left(\frac{1+\lambda}{1+\lambda_{1}}\right)}^{M}
\expp^{-\pi^2 x^2({\lambda_1}^2-\lambda^2)/2}
\end{equation}
After performing the Fourier transform in~(\ref{defform}) to obtain the form factor we arrive at
\begin{equation}
K^{\text{GUE}}(\tau,x,M)=\frac{1}{2} \int_{-1}^{1} \ud \lambda \int_{1}^{\infty} \ud \lambda_1 \;
\delta(\lambda_{1}-\lambda-2\tau) {\left(\frac{1+\lambda}{1+\lambda_1}\right)}^{M}
\expp^{-\pi^{2}x^2({\lambda_{1}}^{2}-\lambda^{2})/2}
\end{equation}
In this equation we have removed a second delta function which does not
contribute due to the restrictions $\tau>0$ and $\lambda_{1}\geq \lambda$.
The remaining delta function provides the relation $2\tau=\lambda_{1}-\lambda$.
In the case $\tau<1$, the domain of integration for $\lambda_1$ is reduced to
$1\leq \lambda_{1}\leq1+2\tau$, and we obtain
\begin{equation}
K^{\text{GUE}}(\tau,x,M)= \frac{1}{2} \int_{1}^{1+2\tau} \ud \lambda_1 \; 
{\left(\frac{1+\lambda_1-2\tau}{1+\lambda_1}\right)}^{M} \expp^{2\pi^{2}x^{2}\tau(\tau-\lambda_1)}
\; , \quad \tau < 1 \; .
\end{equation}
To get a series expansion of this integral, it is useful to remove
the $\tau$ dependence from the limits by a change of variable.  We do so
by using $\lambda_1=1+\tau y_1$, and we also define $B=2 \pi^2 x^2/\kappa$
where $\kappa=1$ and $2$ for the GUE and GOE cases, respectively. The 
integral then becomes
\begin{equation} \label{kgue1}
K^{\text{GUE}}(\tau,x,M)= \frac{\tau}{2}\expp^{-B\tau} \int_{0}^{2} \ud y_1 \; 
{\left(1-\frac{2\tau}{2+\tau y_1}\right)}^{M} \expp^{B\tau^{2}(1-y_1)}
\end{equation}
Now we can expand the integrand in a power series in $\tau$. When we perform
the integration, term by term, and extract an exponential we get the following
expansion up to 9th order (by using MAPLE)
\begin{align} \label{kgue}
K^{\text{GUE}}(\tau,x,M) = \expp^{-(B+M)\tau} & \left[\tau - \frac{M}{6}\tau^{4}
+ \frac{(2B-M)^{2}}{24}\tau^{5} - \frac{M}{15}\tau^{6} - \left(\frac{BM}{15}
- \frac{M^{2}}{20}\right)\tau^{7} \right. \notag \\
&{}- \left(\frac{7M(2B-M)^{2}}{720} + \frac{M}{28}\right)\tau^{8} \notag \\
&{}\left. + \left(\frac{(2B-M)^{4}}{1920} - \frac{BM}{28} +
\frac{401M^{2}}{10080}\right)\tau^{9} + \ldots \right]
\end{align}
The form factor for cases without
parametric variation is obtained by setting $B=0$ (or equivalently $x=0$). It is therefore
\begin{align} \label{kgueopen}
K^{\text{GUE}}(\tau,M) = \expp^{-M \tau}& \left[ \tau - \frac{M\tau^{4}}{6} +
\frac{M^{2}\tau^{5}}{24} - \frac{M\tau^{6}}{15} + \frac{M^{2}\tau^{7}}{20} -
\frac{7M^{3}\tau^{8}}{720} \right. \notag \\
&\left.{} - \frac{M\tau^{8}}{28} + \frac{M^{4}\tau^{9}}{1920} +
\frac{401M^{2}\tau^{9}}{10080} + \ldots \right]
\end{align}

For the GOE case, the correlation function can be found for the case without parametric
variations in \cite{LSSS95}, and for the case with parametric variations in \cite{FSS97}
The parametric correlation function, which we consider here, is expressed by a triple integral
\begin{align} \label{r2goe}
R_2^{\text{GOE}}(\eta,x,M)&= \int_{-1}^{1} \ud \lambda \int_{1}^{\infty} \ud \lambda_1
\int_{1}^{\infty} \ud \lambda_2 \; \cos(\pi\eta(\lambda-\lambda_{1}\lambda_{2}))
\frac{(1-\lambda^{2})(\lambda-\lambda_{1}\lambda_{2})^{2}}{
(2\lambda\lambda_{1}\lambda_{2}-\lambda^{2}-{\lambda_{1}}^{2}-{\lambda_{2}}^{2}+1)^{2}}
\notag \\
& \qquad \qquad \times
{\left(\frac{1+\lambda}{\lambda_{1}+\lambda_{2}}\right)}^{M}\expp^{-\pi^{2}x^2
(2{\lambda_{1}}^{2}{\lambda_{2}}^{2}
-\lambda^{2}-{\lambda_{1}}^{2}-{\lambda_{2}}^{2}+1)/4}
\end{align}
To obtain the parametric form factor we take the Fourier transform and obtain
a sum of two delta functions.  Again, because $\tau$ is positive and
$\lambda_{1}\lambda_{2}\geq \lambda$ only one of the delta function
contributes, and we only include this one in the expression
\begin{align}
K^{\text{GOE}}(\tau,x,M)&= \int_{-1}^{1} \ud \lambda \int_{1}^{\infty} \ud \lambda_1 
\int_{1}^{\infty} \ud \lambda_2 \; \delta(\lambda-\lambda_{1}\lambda_{2}-2\tau)
\frac{(1-\lambda^{2})(\lambda-\lambda_{1}\lambda_{2})^{2}}{
(2\lambda\lambda_{1}\lambda_{2}-\lambda^{2}-{\lambda_{1}}^{2}-{\lambda_{2}}^{2}+1)^{2}}
\notag \\ 
& \qquad \qquad \times
{\left(\frac{1+\lambda}{\lambda_{1}+\lambda_{2}}\right)}^{M}
\expp^{-\pi^{2}x^2(2{\lambda_{1}}^{2}{\lambda_{2}}^{2}-\lambda^{2}-
{\lambda_{1}}^{2}-{\lambda_{2}}^{2}+1)/4}
\end{align}
The delta function yields the relation $\lambda=\lambda_{1}\lambda_{2}-2\tau$. 
As $\tau<1$, our domain of integration for the
other two variables is given by $1\leq \lambda_{1}\leq1+2\tau$ and
$1\leq \lambda_{2}\leq\frac{1+2\tau}{\lambda_{1}}$.  When the integral
over $\lambda$ is performed, we are left with
\begin{align}
K^{\text{GOE}}(\tau,x,M)&=
\int_{1}^{1+2\tau} \ud \lambda_1 \int_{1}^{\frac{1+2\tau}{\lambda_{1}}}  \ud \lambda_2 \;
\frac{4\tau^{2}(1-{\lambda_{1}}^{2}{\lambda_{2}}^{2}+4\tau\lambda_{1}\lambda_{2}-4\tau^{2})
}{(1+{\lambda_{1}}^{2}{\lambda_{2}}^{2}-{\lambda_{1}}^{2}-{\lambda_{2}}^{2}-4\tau^{2})^{2}}
{\left(\frac{1+\lambda_{1}\lambda_{2}-2\tau}{\lambda_{1}+\lambda_{2}}\right)}^{M}\notag \\ 
& \qquad \qquad \times 
\expp^{-\pi^{2}x^{2}(1+{\lambda_{1}}^{2}{\lambda_{2}}^{2}-{\lambda_{1}}^{2}-{\lambda_{2}}^{2}
+4\tau\lambda_{1}\lambda_{2}-4\tau^{2})/4}
\end{align}
In order to compare to semiclassical results this integral is evaluated as a series
in $\tau$.  To do so we first remove the $\tau$ 
dependence from the limits by introducing new variables defined by
$\lambda_1=1+ \tau y_1$ and $\lambda_1 \lambda_2 = 1 + \tau y_2$.
After this change of variables, the integrand is expanded as a series in $\tau$.
We also replace $\pi^{2}x^{2}$ by $B$ (as defined after equation~(\ref{kgue1})),
though we only include the first two orders for clarity 
\begin{align}
K^{\text{GOE}}(\tau,x,M)& = \int_{0}^{2} \ud y_1 \int_{y_{1}}^{2} \ud y_2 \; \left\{
\frac{2-y_{2}}{2(1-y_{1}y_{2}+{y_{1}}^{2})^{2}}\tau + 
\left[\frac{y_{1}(y_{2}-2)(4-y_{1}y_{2}+2{y_{1}}^{2}-{y_{2}}^{2})}{
2(1-y_{1}y_{2}+{y_{1}}^{2})^{3}}\right. \right. \notag \\
& \qquad \qquad \left. \left. + \frac{(2-y_{2})(2+2B+2M-6y_{1}-y_{2})}{
4(1-y_{1}y_{2}+{y_{1}}^{2})^{2}}\right]\tau^{2} + \ldots \right\}
\end{align}
Once the integrals are evaluated we get a series expansion for the parametric
form factor for the time delay.  In order to compare with semiclassical results, we again extract an
exponential term, to get the following result to seventh order
\begin{align} \label{kgoe}
K^{\text{GOE}}(\tau,x,M) = \expp^{-(B+M)\tau} & \left[ 2\tau - 2\tau^{2} - (2B-M-2)\tau^{3}
+ \left(2B - \frac{7M}{3} - \frac{8}{3}\right)\tau^{4} \right. \notag \\
&{} + \left(\frac{5(2B-M)^{2}}{12} - \frac{8B}{3}+\frac{13M}{3} + 4\right)\tau^{5} \notag\\
&- {} \left(\frac{5B^{2}}{3} - \frac{11BM}{3} + \frac{17M^{2}}{12} - 4B + \frac{39M}{5} 
+ \frac{32}{5}\right)\tau^{6} \notag \\
&{}- \left(\frac{41(2B-M)^{3}}{360} - \frac{11B^{2}}{5} + 7BM - \frac{43M^{2}}{12} \right. \notag \\
&\left.\left.{}+ \frac{32B}{5} - \frac{212M}{15} - \frac{32}{3}\right)\tau^{7} + \ldots \right]
\end{align}
By setting $B=0$ we get the form factor for the case without parametric correlations
\begin{align} \label{kgoeopen}
K^{\text{GOE}}(\tau,M) = \expp^{-M\tau} & \left[2\tau - 2\tau^{2} + (M+2)\tau^{3} - 
\left(\frac{7M}{3} + \frac{8}{3}\right)\tau^{4}\right. \notag \\
&{}+ \left(\frac{5M^{2}}{12} + \frac{13M}{3} + 4\right)\tau^{5} - \left(\frac{17M^{2}}{12} 
+ \frac{39M}{5} + \frac{32}{5}\right)\tau^{6} \notag \\
&\left.{}+ \left(\frac{41M^{3}}{360} + \frac{43M^{2}}{12} + \frac{212M}{15} + 
\frac{32}{3}\right)\tau^{7} + \ldots \right]
\end{align}

Finally, we include the dependence on a further parameter $y$ that breaks time-reversal symmetry
and thus leads to a GOE-GUE transition as $y$ is varied from zero to infinity. The RMT result
can again be found in reference \cite{FSS97}, the equation (2) in this article includes even
more general cases. For our case, the correlation function has the form of the integral (\ref{r2goe})
with an additional factor in the integrand that is given by
\begin{align}
G(y) & = \exp\left( \frac{y}{8}(\lambda^2 + 1 - 2 \lambda_2^2) \right)
\notag \\ & \qquad \times
\left[(1-\lambda^2) (1 + \frac{y R}{4}) \cosh \alpha
+ (\lambda_2^2 - \lambda_1^2) \sinh \alpha + 
\frac{y R}{4} (2 \lambda_2^2 + \lambda^2 - 1) \sinh \alpha \right]
\end{align}
where
\begin{equation}
R = \lambda_1^2 + \lambda_2^2 + \lambda^2 - 2 \lambda \lambda_1 \lambda_2 - 1 \; , \quad
\text{and} \quad \alpha = \frac{y}{8} (1 - \lambda^2)
\end{equation}
and we choose the parameters $y_1$ and $y_2$ in \cite{FSS97} as $y_1^2 = y_2^2 = y/8$.
The steps to get the form factor are exactly the same as for the GOE case and we
give only the final result
\begin{align} \label{rmtall}
K(\tau,x,y,M) 
= & \expp^{-(B+M) \tau} \left[
2 \tau - \left( 2 + y \right) \tau^2
+ \left( 2 + 2 y + \frac{1}{2} y^2 - 2 B + M  \right) \tau^3 
- \left( \frac{8}{3} + 2 y 
\right. \right. \notag \\ & \left.
+ y^2 + \frac{1}{6} y^3 - 2 B - B y 
+ \frac{7}{3} M + \frac{1}{2} M y \right) \tau^4
+ \left(4 + 2 y + \frac{7}{6} y^2 + \frac{1}{3} y^3 + \frac{1}{24} y^4
\right. \notag \\ & \left. \left.
- \frac{8}{3} B + \frac{13}{3} M + \frac{5}{12} (2 B - M)^2 
- \frac{4}{3} (B-M) y  - \frac{1}{6} (2 B - M) y^2
\right) \tau^5 
+ \ldots \right]
\end{align}

Unfortunately, one cannot see the transition from GOE to GUE in this 
formula. As will become clear from the semiclassical calculation,
there are exponential terms of the form $\exp(- y \tau)$ which describe
this transition, but these have been expanded when we expanded
the integrand in powers of $\tau$. We tried to replace $y$ by
$\tilde{y}/\tau$ in order to extract these exponential terms, but
then we could not bring MAPLE to perform the integrals. However
the equation (\ref{rmtall}) can be compared to the semiclassical
result if we expand the exponentials there as well.

\section{Semiclassical approximation}\label{diagopen}

In this section we derive a semiclassical expression for the
form factor. The oscillatory part of the density in (\ref{reldens})
has exactly the same semiclassical form as in Gutzwiller's
trace formula for the density of states \cite{BB74,Gut71}
\begin{equation} \label{gutz}
d^\text{osc}(E,X) \approx \frac{1}{\pi \hbar} \Re
\sum_\gamma A_\gamma \exp\left( \frac{\ci}{\hbar} S_\gamma \right)
\; , \quad
\text{where} \quad A_\gamma = \frac{T_\gamma}{R_\gamma
\sqrt{|\det(M_\gamma - 1)|}} \expp^{-\ci \pi \mu_\gamma/2} \; .
\end{equation}
The sum runs over all periodic orbits in the repeller with period
$T_\gamma$, repetition number $R_\gamma$, stability matrix
$M_\gamma$ and Maslov index $\mu_\gamma$.

In order to evaluate the two-point correlation function of the
form factor we expand the action in first order in the energy
difference and the parameter difference
\begin{equation} \label{act}
S_\gamma \left(E \pm \frac{\eta}{2 \bar{d}} \pm \frac{x \rho}{2 \sigma}
,X \pm \frac{x}{2 \sigma} \right)
\approx S_\gamma(E,X) \pm T_\gamma(E,X) \frac{\eta}{2 \bar{d}}
\pm Q_\gamma(E,X) \frac{x}{2 \sigma}
\end{equation}
where 
\begin{equation}
Q_\gamma=\left. \frac{\partial S_\gamma}{\partial X} \right|_{\tilde{E}}    
=  \rho \frac{\partial S_\gamma}{\partial E} 
 + \frac{\partial S_\gamma}{\partial X}
\end{equation}
is the parametric velocity. In this and the next chapter we consider systems
without a parameter that breaks time reversal symmetry. The case with
such a parameter will be discussed in section~\ref{mag}.

After inserting (\ref{gutz}) and (\ref{act}) into (\ref{r2}) and (\ref{defform})
and evaluating the integral to leading semiclassical order, we arrive at
\begin{equation} \label{ksemi}
K(\tau,x) = \frac{1}{T_H} \left\langle \sum_{\gamma,\gamma'} 
A_\gamma A_{\gamma'}^* \expp^{\frac{ \ci(S_\gamma-S_{\gamma'})}{\hbar}}
\expp^{\frac{\ci x (Q_\gamma + Q_{\gamma'})}{2 \sigma \hbar}}
\delta \left( T - \frac{T_\gamma + T_{\gamma'}}{2} \right) \right\rangle
\end{equation}
where $T_H = 2 \pi \hbar \bar{d}(E)$ is the Heisenberg time, and $\tau=T/T_H$.
Terms which have a sum of the actions in the exponent have been neglected,
because they are averaged away. Equation (\ref{ksemi}) has exactly the same form
as for the parametric form factor for closed systems \cite{KS06}. The difference
is that the sum is over the periodic orbits in an open system which obey
a different sum rule than the periodic orbits in a closed system. 

The diagonal approximation involves pairs of orbits that are either
identical or related by time reversal. It has the form 
\begin{equation}
K(\tau,x,M) = \frac{\kappa}{T_H} \left\langle \sum_\gamma 
|A_\gamma|^2 \expp^{\frac{\ci x Q_\gamma}{\sigma \hbar}}
\delta(T - T_\gamma) \right\rangle \; ,
\end{equation}
where $\kappa$ is 2 if the system has time reversal symmetry
and 1 if it does not. 

An important input in the following semiclassical calculation
is the distribution of the parametric velocities $Q_\gamma$ in
the limit of very long periodic orbits. The $Q_\gamma$ have a
mean value of zero and a variance proportional to their period \cite{GSBSWZ91}
\begin{equation} \label{qmean}
\langle Q_\gamma \rangle = 0 \; , \quad \langle Q_\gamma^2 \rangle \sim a T \; ,
\quad T \rightarrow \infty
\end{equation}
where the averages are performed over trajectories with period
around $T$. 

The main assumption that we will use in the following
is that the $Q_\gamma$ have a Gaussian distribution for
long periodic orbits. This has been motivated e.g. in \cite{OLM98}.
In closed systems the proportionality factor $a$ in (\ref{qmean})
is related to the variance of the level velocities by \cite{EFKAMM95,LS99}
\begin{equation} \label{semisigma}
\sigma^2 \sim \frac{a \kappa \bar{d}}{2 \pi \hbar} \; .
\end{equation}
We will see that we obtain the universal form factor when we
use this relation for the scaling factor $\sigma$ in (\ref{r2})
in open systems as well.

With the Gaussian assumption for the distribution of the parametric
velocities and (\ref{semisigma}) the average over the $Q_\gamma$
can be performed, assuming that it can be done independently, and
we obtain
\begin{equation} \label{qaverage}
\langle \expp^{\frac{\ci x Q_{\gamma}}{\sigma \hbar}} \rangle
= \expp^{-\frac{x^2 a T}{2 \sigma^2 \hbar^2}} = \expp^{-B T/T_H}
\end{equation}
where $B = 2 \pi^2 x^2/\kappa$ has been introduced after equation
(\ref{kgue1}). The remaining sum over periodic orbits can be evaluated 
with a modification of the Hannay-Ozorio de Almeida sum rule \cite{HO84}.
This modification takes into account the probability that classical
trajectories escape in open systems.
The classical escape rate, or the inverse of the mean time spent in the
cavity, can be expressed as $\mu=M/T_{H}$ (see e.g. \cite{LV04}), and the
sum rule takes the form
\cite{CE91}
\begin{equation} \label{summodi}
\sum_\gamma |A_\gamma|^2 \delta(T - T_\gamma) \approx T \expp^{-M T/T_H} \; .
\end{equation}
We find that the diagonal approximation is given by
\begin{equation} 
K_d(\tau,x,M) = \kappa \tau \expp^{-(B+M) \tau}
\end{equation}
in agreement with the first term in the expansion of the
random matrix results, (\ref{kgue}) and (\ref{kgoe}). Previous
evaluations of the diagonal approximation of the form factor
for the time delay, respectively its Fourier transform, were
performed in \cite{Eck93,VOL98}. The second reference also
treats time reversal symmetry breaking.

\section{Off-diagonal contributions}

The correlations between different periodic orbits
in open systems that are responsible for the off-diagonal
terms in the expansion of the form factor have the same origin
as in closed systems \cite{SR01,MHBHA04,MHBHA05}. The general
formalism for evaluating the correlations between orbits for
arbitrary order has been developed in \cite{MHBHA04,MHBHA05,Mue05}.

Long periodic orbits have many close self-encounters in which
two or more stretches of an orbit are almost identical,
possibly up to time reversal. These orbit stretches are
connected by long parts of the orbit which are called ``loops''.
There exist other periodic orbits which are almost identical
along the loops, but they differ in the way in which the loops
are connected in the encounter regions. As a consequence the
actions, periods and stabilities of these orbits are strongly
correlated. One can classify correlated orbit pairs according
to ``structures'' that are characterized by the number of
encounter regions $V$ in which the loops are connected in a
different way, the number of involved orbit stretches $l_\alpha$
in each encounter region $\alpha$, and the way in which the loops
are connected by these stretches. Structures can be put in a
one-to-one relation with permutation matrices that describe
the reconnections of the loops in the encounter region.
It is a combinatorial problem to find all possible structures
that describe correlations between periodic orbits.

The number and types of encounter regions are specified by a
vector $\v$ whose $l$-th component, $v_l$, denotes the number of
encounter regions with $l$ stretches. The total number of
orbit stretches is $L$. Hence
\begin{equation}
V = \sum_{l \ge 2} v_l \; , \qquad L = \sum_\alpha l_\alpha = \sum_{l \ge 2} l v_l \; .
\end{equation} 
In general there are many structures with the same vector $\v$,
because, for example, there are many different ways in which the loops can
be reconnected in the encounter regions, and the number of
structures with the same vector $\v$ is denoted by ${\cal N}(\v)$.

The summation over all orbit pairs with the same structure
can be performed by using that long periodic orbits are
uniformly distributed over the surface of constant energy
in closed systems. In open systems one has take additionally
into account that trajectories can escape to infinity. 
One combines the contributions from all structures
with the same vector $\v$, and the semiclassical form factor
is then obtained by summing over all $\v$ and adding the
diagonal approximation 
\begin{equation}
K^{\text{sc}}(\tau) = \kappa \tau + \sum_{\v} K_{\v}(\tau)
\end{equation}
where $K_{\v}(\tau)$ is the contribution to the form factor
from all structures with the same $\v$.

The derivation of $K_{\v}(\tau)$ for the {\em spectral}
form factor can be found in \cite{MHBHA05,Mue05}. We state here
only those details that are relevant for the following calculations.
The semiclassical expression for $K_{\v}(\tau)$ can be expressed as
\begin{equation} \label{sumv}
K_{\v}(\tau) = \frac{1}{T_H} \sum_{(\gamma,\gamma')}^{\text{fixed} \, \v} 
|A_\gamma|^2 e^{\ci \Delta S_\gamma/\hbar} 
\delta(T - T_\gamma) = N(\v) \kappa \tau \int d^{L-V} s \; d^{L-V} u \;
\frac{w_T(\s,\u)}{L} \expp^{\ci \s \u / \hbar} \; .
\end{equation}
Here the components of the vectors $\s$ and $\u$ are coordinates along the
stable and unstable manifolds in Poincar\'e sections in
the $V$ encounter regions, that describe the relative positions
of the orbit stretches. In (\ref{sumv}) the amplitudes and periods
of the two correlated orbits are set equal. $w_T(\s,\u)$ is the
density of self-encounters for a given structure and separation
coordinates $\s$ and $\u$. For long orbits it is given asymptotically by
\begin{equation} \label{defw}
\frac{w_T(\s,\u)}{L} = \frac{T (T - \sum_\alpha l_\alpha t_{\text{enc}}^\alpha)^{L-1}}{
L! \Omega^{L-V} \prod_\alpha t_{\text{enc}}^\alpha } \; .
\end{equation}
Here $\alpha$ labels the $V$ different encounters, each being a
$l_{\alpha}$-encounter, and $t_{\text{enc}}^\alpha$ is the time for the traversal
of the encounter region $\alpha$. The traversal time is a function of $\s$ and $\u$.

The only property that is needed for an evaluation of the integral in (\ref{sumv})
is the semiclassical relation
\begin{equation} \label{intenc}
\int \prod_j \ud s_{\alpha j} \; \ud u_{\alpha j} \; (t_{\text{enc}}^\alpha)^k 
\expp^{\ci \sum_j s_{\alpha j} u_{\alpha j} / \hbar} \approx
\begin{cases} 0 & \text{if} \; k=-1 \; \text{or}  \; k \ge 1 \\
              (2 \pi \hbar)^{l_\alpha-1} & \text{if} \; k=0 \end{cases} \; .
\end{equation}
This considerably simplifies the evaluation of the contribution $K_{\v}(\tau)$,
because after an expansion of the numerator of (\ref{defw}) the only terms that
survive are the ones that contain a product of all encounter times
$t_{\text{enc}}^\alpha$ which is cancelled by the denominator. In this
way the contribution $K_{\v}(\tau)$ was evaluated for the spectral
form factor in \cite{MHBHA04,MHBHA05}. By deriving a recursion relation
involving the numbers ${\cal N}(\v)$ the semiclassical expansion could
be summed up, and it was shown that it agrees with form factor of RMT
for $\tau < 1$.

\subsection{Off-diagonal terms without parametric correlations}

Let us now come back to the form factor for the time delay.
We consider first the case without parametric correlations, i.e. $x=0$.
We have to evaluate
\begin{equation}
K_{\v}(\tau,M) = \frac{1}{T_H} \sum_{(\gamma,\gamma')}^{\text{fixed} \, \v} 
|A_\gamma|^2 \expp^{\frac{ \ci(S_\gamma-S_{\gamma'})}{\hbar}}
\delta \left( T - T_\gamma\right)
\end{equation}
where the sum is over all correlated pairs of periodic orbits in open
systems with the same number and type of encounter regions, specified by $\v$.
The density $w_T(\s,\u)$ in (\ref{sumv}) has to be modified by the
survival probability of long classical trajectories in open systems,
similarly as was done in the sum rule for the diagonal approximation
(\ref{summodi}). If there would be no systematic correlations between
different parts of an orbit then this survival property would be
$\exp(-\mu T)$. However the density $w_T(\s,\u)$ describes orbits 
that all have the same number and type of encounter regions.
In an encounter region there are different stretches of an orbit
which are almost identical. If an orbit does not escape during
the traversal of one stretch in an encounter region then it 
has a negligible probability of escaping during the traversal
of the other stretches in the same encounter region. Hence
for each encounter region the survival time $T$ should be
reduced by $(l_\alpha - 1) t_{\text{enc}}^\alpha$. The total
survival probability is then
\begin{equation} \label{qmaver2}
\expp^{-\mu T} \expp^{ \mu \sum_\alpha (l_\alpha-1) t_{\text{enc}}^\alpha}
\end{equation}
This is analogous to the survival probability for correlated
pairs of open trajectories that contribute to the conductance \cite{HMBH06}.

To summarize, the contribution of all orbit pairs with encounter regions
described by $\v$ is
\begin{equation} \label{sumvp}
K_{\v,M}(\tau) = \frac{1}{T_H} \sum_{(\gamma,\gamma')}^{\text{fixed} \, \v} 
|A_\gamma|^2 e^{\ci \Delta S_\gamma/\hbar} 
\delta(T_H \tau - T_\gamma) = N(\v) \kappa \tau \int d^{L-V} s \; d^{L-V} u \;
\frac{w_T(\s,\u)}{L} \expp^{\ci \s \u / \hbar}
\end{equation}
where the density of self-encounters is now
\begin{equation} \label{defw2}
\frac{w_{T}(s,u)}{L}= \frac{\expp^{-MT/T_H}T(T-\sum_{\alpha}l_{\alpha}t_{enc}^{\alpha})^{L-1}
\prod_{\alpha}\expp^{M(l_\alpha-1)t_{enc}^{\alpha}/T_H}}{L!\Omega^{L-V}
\prod_{\alpha}t_{enc}^{\alpha}} \; ,
\end{equation}
which has been obtained by multiplying (\ref{defw}) by (\ref{qmaver2}). 
Again because of (\ref{intenc}) the only terms that contribute in the semiclassical
limit are those where the encounter times in the numerator and denominator cancel
exactly. As a first step we can expand the exponentials as a power series up to
first order
\begin{equation} \label{ww1}
\frac{w_{T}(s,u)}{L} \Longrightarrow 
\frac{\expp^{-M T/T_H}T(T-\sum_{\alpha}l_{\alpha}t_{enc}^{\alpha})^{L-1}
\prod_{\alpha}(1+(l_{\alpha}-1) M t_{enc}^{\alpha}/T_H)}{L!\Omega^{L-V}
\prod_{\alpha}t_{enc}^{\alpha}}
\end{equation}
The method to obtain $K_{\v}(\tau,M)$ from the expression (\ref{ww1}) is
straighforward. For a given vector $\v$ (which specifies the set of the
$l_\alpha$) the numerator is expanded in all encounter times
$t_{enc}^{\alpha}$ and only those terms are kept for which the product
of the encounter times is exactly cancelled by that in the denominator.
This is conveniently done with a computer program. $K_{\v}(\tau,M)$
follows by multiplying the result with ${\cal N}(\v) \kappa \tau (2 \pi \hbar)^{L-V}$,
because of (\ref{sumvp}) and (\ref{intenc}). Note that in order to
obtain a product of the $V$ different encounter times in the numerator
we can take $r$ of them from the product over $\alpha$ and $V-r$ of them
from the bracket with the power $L-1$, where $0 \leq r \leq V$. This 
gives a term to the form factor which is proportional to $\expp^{-M \tau} \tau^{L-V+1+r} M^r$.

Table~\ref{orbittableopen} shows the result for the different types of encounters
with $L-V \leq 4$. The vectors $\v$ are represented in the form
$(2)^{v_2}(3)^{v_3} \ldots$ and the horizontal lines separate vectors $\v$ with
different value of $L-V$. The numbers $N(\v)$ can be calculated by combinatorial
methods \cite{Mue05}.

\begin{table}[htb]
\centering
\begin{tabular}{|c|c|c|c|c|c|}
\hline
$\v$ & $L$ & $V$ & $K_{\v}(\tau,M)/(\kappa N(\v))$ & $N(\v)$, no TRS& $N(\v)$, TRS\\
\hline
$(2)^{1}$&2&1&$\expp^{-M\tau}\left(-\tau^{2}+\frac{M\tau^{3}}{2}\right)$&-&1\\
\hline
$(2)^{2}$&4&2&$\expp^{-M\tau}\left(\tau^{3}-\frac{M\tau^{4}}{2}+\frac{M^{2}\tau^{5}}{24}\right)$&1&5\\
$(3)^1$&3&1&$\expp^{-M\tau}\left(-\tau^{3}+\frac{M\tau^{4}}{3}\right)$&1&4\\
\hline
$(2)^{3}$&6&3&$\expp^{-M\tau}\left(-\frac{2\tau^{4}}{3}+\frac{M\tau^{5}}{3}-\frac{M^{2}\tau^{6}}{24}+\frac{M^{3}\tau^{7}}{720}\right)$&-&41\\
$(2)^{1}(3)^1$&5&2&$\expp^{-M\tau}\left(\frac{3\tau^{4}}{5}-\frac{7M\tau^{5}}{30}+\frac{M^{2}\tau^{6}}{60}\right)$&-&60\\
$(4)^{1}$&4&1&$\expp^{-M\tau}\left(-\frac{\tau^{4}}{2}+\frac{M\tau^{5}}{8}\right)$&-&20\\
\hline
$(2)^{4}$&8&4&$\expp^{-M\tau}\left(\frac{\tau^{5}}{3}-\frac{M\tau^{6}}{6}+\frac{M^{2}\tau^{7}}{40}-\frac{M^{3}\tau^{8}}{720}+\frac{M^{4}\tau^{9}}{40320}\right)$&21&509\\
$(2)^{2}(3)^1$&7&3&$\expp^{-M\tau}\left(-\frac{2\tau^{5}}{7}+\frac{5M\tau^{6}}{42}-\frac{11M^{2}\tau^{7}}{840}+\frac{M^{3}\tau^{8}}{2520}\right)$&49&1092\\
$(2)^{1}(4)^1$&6&2&$\expp^{-M\tau}\left(\frac{2\tau^{5}}{9}-\frac{5M\tau^{6}}{72}+\frac{M^{2}\tau^{7}}{240}\right)$&24&504\\
$(3)^{2}$&6&2&$\expp^{-M\tau}\left(\frac{\tau^{5}}{4}-\frac{M\tau^{6}}{12}+\frac{M^{2}\tau^{7}}{180}\right)$&12&228\\
$(5)^{1}$&5&1&$\expp^{-M\tau}\left(-\frac{\tau^{5}}{6}+\frac{M\tau^{6}}{30}\right)$&8&148\\
\hline
\end{tabular}
\caption{Contribution of different types of orbit pairs to the form factor for the time delay}
\label{orbittableopen}
\end{table}

To find the total contribution to the form factor we now multiply the middle column 
that contains $K_{\v}(\tau,M)/(\kappa N(\v))$ by $\kappa$ and $N(\v)$, add the diagonal approximation
and sum over different $\v$. If we do that for all orbits pairs with $L-V\leq8$
for the case without time reversal symmetry ($\kappa=1$), we obtain the expansion
for the form factor in $\tau$ up to 9th order
\begin{align} \label{gue_m}
K(\tau,M)=\expp^{-M\tau} & \left[\tau - \frac{M\tau^{4}}{6} + \frac{M^{2}\tau^{5}}{24}
- \frac{M\tau^{6}}{15} + \frac{M^{2}\tau^{7}}{20} \right. \notag\\
& \left. {}- \frac{7M^{3}\tau^{8}}{720} - \frac{M\tau^{8}}{28}
+ \frac{M^{4}\tau^{9}}{1920} + \frac{401M^{2}\tau^{9}}{10080} + \ldots \right]
\end{align}
This agrees with the terms of the expansion (\ref{kgueopen}) in the section
on RMT.

For systems with time reversal symmetry ($\kappa = 2$) we sum over all contributions
with $L-V\leq6$, and obtain the expansion of the open form factor in $\tau$ up
to 7th order.
\begin{align} \label{goe_m}
K(\tau,M) = \expp^{-M\tau} & \left[2\tau - 2\tau^{2} + (M+2)\tau^{3}
- \left(\frac{7M}{3} + \frac{8}{3}\right)\tau^{4}\right. \notag \\
&{}+ \left(\frac{5M^{2}}{12} + \frac{13M}{3} + 4\right)\tau^{5}
- \left(\frac{17M^{2}}{12} + \frac{39M}{5} + \frac{32}{5}\right)\tau^{6} \notag \\
&\left.{}+ \left(\frac{41M^{3}}{360} + \frac{43M^{2}}{12}
+ \frac{212M}{15} + \frac{32}{3}\right)\tau^{7} + \ldots \right]
\end{align}
This agrees with the expansion (\ref{kgoeopen}) of the GOE result.

\subsection{Off-diagonal terms with parametric correlations}

Let us now consider the parametric form factor $K(\tau,x,M)$. We
have to evaluate
\begin{equation}
K_{\v}(\tau,x,M) = \frac{1}{T_H} \sum_{(\gamma,\gamma')}^{\text{fixed} \, \v} 
|A_\gamma|^2 \expp^{\frac{ \ci(S_\gamma-S_{\gamma'})}{\hbar}}
\expp^{\frac{\ci x Q_\gamma}{\sigma \hbar}}
\delta \left( T - T_\gamma\right)
\end{equation}
It now contains an additional term involving the parametric velocities $Q_\gamma$.
As for the diagonal approximation we assume that the average over the $Q_\gamma$
can be performed independently from the actions and amplitudes of the orbits.
However, we have to look again at the systematic correlations between different
parts of the {\em same} periodic orbit that are implied by fixing the vector $\v$
\cite{KS06}. In the encounter regions the different stretches of an orbit are
almost identical, and the same holds for the changes of the action along these
stretches when an external parameter is varied. Hence we should consider the
average over the parametric velocities for the loops and encounter regions separately.
The contribution from a loop is
\begin{equation} 
\left\langle\expp^{\frac{\ci xQ^{loop}_{\gamma}}{\sigma\hbar}}\right\rangle=
\expp^{-BT_{\text{loop}}/T_H}
\end{equation}
while the contribution from the $l$ orbit stretches in an $l$-encounter region is
\begin{equation} 
\left\langle\expp^{\frac{\ci xlQ^{enc}_{\gamma}}{\sigma\hbar}}\right\rangle
=\expp^{-B l^2 t_{enc}/T_H} \; .
\end{equation}
This means that the average over the parametric velocities is now given by
\begin{equation} \label{qaver2}
\left\langle\expp^{\frac{\ci xQ_{\gamma}}{\sigma\hbar}}\right\rangle =
\expp^{-B (T - \sum_\alpha l_\alpha t_{\text{enc}}^\alpha) /T_H}
\expp^{-B \sum_\alpha l_\alpha^2 t_{\text{enc}}^\alpha /T_H}
\end{equation}

The contribution of all orbit pairs with encounter regions described by $\v$ then
have the form
\begin{equation} \label{sumvp2}
K_{\v}(\tau,x,M) = \frac{1}{T_H} \sum_{(\gamma,\gamma')}^{\text{fixed} \, \v} 
|A_\gamma|^2 e^{\ci \Delta S_\gamma/\hbar} 
\delta(T - T_\gamma) = N(\v) \kappa \tau \int d^{L-V} s \; d^{L-V} u \;
\frac{z_T(\s,\u)}{L} \expp^{\ci \s \u / \hbar} \; .
\end{equation}
were $z_T(\s,\u)$ is the product of the density of self-encounters $w_T(\s,\u)$ in open systems
in (\ref{defw2}) and the average over the phase factors in (\ref{qaver2})
\begin{equation} 
\frac{z_{T}(s,u)}{L}=  \frac{w_{T}(s,u)}{L}
\left\langle\expp^{\frac{\ci xQ_{\gamma}}{\sigma\hbar}}\right\rangle 
=\frac{\expp^{-(B+M)T/T_H}T(T-\sum_{\alpha}l_{\alpha}t_{enc}^{\alpha})^{L-1}
\prod_{\alpha}\expp^{-(Bl_\alpha-M)(l_\alpha-1)t_{enc}^{\alpha}/T_H}}{L!\Omega^{L-V}
\prod_{\alpha}t_{enc}^{\alpha}} \; .
\end{equation}
As before, because of (\ref{intenc}), the only terms that contribute in the
semiclassical limit are those where the encounter times in the numerator and
denominator cancel exactly. As a first step we can expand the exponentials as
a power series up to first order
\begin{equation} 
\frac{z_{T}(s,u)}{L} \Longrightarrow \frac{\expp^{-(B+M) T/T_H}
T (T-\sum_{\alpha}l_{\alpha}t_{enc}^{\alpha})^{L-1}
\prod_{\alpha}(1-(Bl_\alpha-M)(l_{\alpha}-1) t_{enc}^{\alpha}/T_H)}{L!\Omega^{L-V}
\prod_{\alpha}t_{enc}^{\alpha}}
\end{equation}
We proceed as in the previous subsection. 
For a given vector $\v$ the numerator is expanded in all encounter times
$t_{enc}^{\alpha}$ and only those terms are kept for which the product
of the encounter times is exactly cancelled by that in the denominator.
This is done with a computer program. We multiply the result 
with ${\cal N}(\v) \kappa \tau (2 \pi \hbar)^{L-V}$ and obtain $K_{\v}(\tau,x,M)$.
Note that in order to obtain a product of the $V$ different encounter times
in the numerator we can take $r$ of them from the product over $\alpha$
and $V-r$ of them from the bracket with the power $L-1$, where $0 \leq r \leq V$.
This gives terms to the form factor which are proportional to
$\expp^{-(B+M) \tau} \tau^{L-V+1+r} B^s M^{r-s}$, where $0 \leq s \leq r$.

The contribution of orbits for the different types of encounters
are shown for orbit pairs with $L-V \leq 4$ in table~\ref{orbittableparaopen}.
The double horizontal lines separate vectors $\v$ with different value of $L-V$.

\begin{table}[htb]
\centering
\begin{tabular}{|c|c|c|c|c|c|}
\hline
\multirow{2}{*}{$\v$} &\multirow{2}{*}{$L$} & \multirow{2}{*}{$V$} & 
\multirow{2}{*}{$K_{\v}(\tau,x,M)\expp^{(B + M)\tau}/(\kappa N(\v))$} & $N(\v)$,& $N(\v)$\\
& & & & no TRS & TRS \\ 
\hline \hline
$(2)^{1}$&2&1&$-\tau^{2} - \frac{1}{2} \left(2 B - M \right)\tau^{3}$&-&1\\
\hline \hline
$(2)^{2}$&4&2&$\tau^{3}+
\frac{1}{2} \left(2 B - M\right)\tau^{4} + \frac{1}{24} (2 B - M)^2 \tau^{5} $&1&5\\
\hline 
$(3)^{1}$&3&1&$-\tau^{3} - \frac{1}{3} (3 B - M) \tau^{4}$&1&4\\
\hline \hline
\multirow{2}{*}{$(2)^{3}$}&\multirow{2}{*}{6}&\multirow{2}{*}{3}&
$-\frac{2}{3}\tau^{4} - \frac{1}{3} (2B -M) \tau^{5}
- \frac{1}{24} (2 B - M)^2 \tau^{6}$ 
&\multirow{2}{*}{-}&\multirow{2}{*}{41}\\
& & & $- \frac{1}{720} (2 B - M)^3 \tau^{7}$ & & \\
\hline
$(2)^{1}(3)^{1}$&5&2&$\frac{3}{5}\tau^{4} + \left(\frac{3B}{5} 
- \frac{7M}{30}\right)\tau^{5} + \left(\frac{B^{2}}{10} - \frac{BM}{12} 
+ \frac{M^{2}}{60}\right)\tau^{6}$&-&60\\
\hline
$(4)^{1}$&4&1&$-\frac{1}{2}\tau^{4} - \frac{1}{8} (4 B - M) \tau^{5}$&-&20\\
\hline \hline
\multirow{3}{*}{$(2)^{4}$}&\multirow{3}{*}{8}&\multirow{3}{*}{4}&
$\frac{1}{3}\tau^{5} + \frac{1}{6} ( 2 B - M) \tau^{6} 
+ \frac{1}{40} (2 B - M)^2 \tau^{7}$ 
&\multirow{2}{*}{21}&\multirow{2}{*}{509}\\
& & & $+ \frac{1}{720} (2 B - M)^3 \tau^{8}
+ \frac{1}{40320} (2 B - M)^4 \tau^{9}$ & & \\
\hline
\multirow{2}{*}{$(2)^{2}(3)^{1}$}&\multirow{2}{*}{7}&\multirow{2}{*}{3}&
$-\frac{2}{7}\tau^{5} - \left(\frac{2B}{7} - \frac{5M}{42}\right)\tau^{6} 
- \left(\frac{B^{2}}{14} - \frac{13BM}{210} + \frac{11M^{2}}{840}\right)\tau^{7}$ 
&\multirow{2}{*}{49}&\multirow{2}{*}{1092}\\
& & & $-\left(\frac{B^{3}}{210} - \frac{2B^{2}M}{315} + \frac{BM^{2}}{360} 
- \frac{M^{3}}{2520}\right)\tau^{8}$ & & \\
\hline
$(2)^{1}(4)^{1}$&6&2&$\frac{2}{9}\tau^{5} + \left(\frac{2B}{9} - \frac{5M}{72}\right)\tau^{6} + \left(\frac{B^{2}}{30} - \frac{BM}{40} + \frac{M^{2}}{240}\right)\tau^{7}$&24&504\\
\hline
$(3)^{2}$&6&2&$\frac{1}{4}\tau^{5} + \frac{1}{12} (3 B - M) \tau^{6} 
+ \frac{1}{180} (3 B - M)^2 \tau^{7}$&12&228\\
\hline
$(5)^{1}$&5&1&$-\frac{1}{6}\tau^{5} - \frac{1}{30} (5 B - M) \tau^{6}$&8&148\\
\hline
\end{tabular}
\caption{Contribution of different types of orbit pairs to the parametric form factor
for the time delay}
\label{orbittableparaopen}
\end{table}

To obtain the total contribution to the form factor we multiply the middle column 
that contains $K_{\v}(\tau,x,M)/(\kappa N(\v))$ by $\kappa$ and $N(\v)$, 
add the diagonal approximation
and sum over different $\v$. If we do that for all orbits pairs with $L-V\leq8$
for the case without time reversal symmetry ($\kappa=1$), we obtain the expansion
for the form factor in $\tau$ up to 9th order
\begin{align}
K(\tau,x,M) = \expp^{-(B+M)\tau} & \left[\tau - \frac{M}{6}\tau^{4} 
+ \frac{(2B-M)^{2}}{24}\tau^{5} - \frac{M}{15}\tau^{6} - \left(\frac{BM}{15} 
- \frac{M^{2}}{20}\right)\tau^{7} \right. \notag \\
&{}- \left(\frac{7M(2B-M)^{2}}{720} + \frac{M}{28}\right)\tau^{8} \notag \\
&{}\left. + \left(\frac{(2B-M)^{4}}{1920} - \frac{BM}{28} 
+ \frac{401M^{2}}{10080}\right)\tau^{9} + \ldots \right]
\end{align}
This agrees with the expansion (\ref{kgue}) in the chapter on RMT. If we remove the
parametric correlations by setting $B=0$ we obtain the result (\ref{gue_m}). 
If we close the system by setting $M=0$ we get the expansion of the
parametric form factor \cite{KS06}.

For systems with time reversal symmetry ($\kappa = 2$) we sum over all contributions
with $L-V\leq6$, and obtain the expansion of the form factor in $\tau$ up
to 7th order.
\begin{align}
K(\tau,x,M) = \expp^{-(B+M)\tau} & \left[ 2\tau - 2\tau^{2} - (2B-M-2)\tau^{3}
+ \left(2B - \frac{7M}{3} - \frac{8}{3}\right)\tau^{4} \right. \notag \\
&{} + \left(\frac{5(2B-M)^{2}}{12} - \frac{8B}{3}+\frac{13M}{3} + 4\right)\tau^{5} \notag\\
&- {} \left(\frac{5B^{2}}{3} - \frac{11BM}{3} + \frac{17M^{2}}{12} - 4B
+ \frac{39M}{5} + \frac{32}{5}\right)\tau^{6} \notag \\
&{}- \left(\frac{41(2B-M)^{3}}{360} - \frac{11B^{2}}{5} + 7BM - \frac{43M^{2}}{12} \right. \notag \\
&\left.\left.{}+ \frac{32B}{5} - \frac{212M}{15} - \frac{32}{3}\right)\tau^{7} + \ldots \right]
\end{align}
This agrees with the expansion of the RMT result (\ref{kgoe}). For $B=0$ it
agrees with the result (\ref{goe_m}) in the previous section, and for $M=0$
with the parametric form factor \cite{KS06}.

\section{GOE-GUE transition} \label{mag}

We introduce now second parameter, a magnetic field $Y$, that takes a time reversal
invariant system (at $Y=0$) and breaks the symmetry as it is increased ($Y \to \infty$).
The semiclassical evaluation of the form factor for closed systems with a time-reversal
symmetry breaking magnetic field has been carried out in \cite{BGOS95,NS06,NBMSHH06}, and
this calculation can be transferred to the parametric form factor for the time delay.

The magnetic field adds a term $\theta_{\gamma}(B) = \int_{\gamma} \A \ud \q$ 
to the action, where $\A$ is the vector potential of the magnetic field that is
assumed to be constant and perpendicular to our two-dimensional system.
The semiclassical form factor is
\begin{equation} \label{ksemi2}
K(\tau,x,y,M) = \frac{1}{T_H} \left\langle \sum_{\gamma,\gamma'} 
A_\gamma A_{\gamma'}^* \expp^{\frac{ \ci(S_\gamma-S_{\gamma'})}{\hbar}}
\expp^{\frac{ \ci(\theta_\gamma-\theta_{\gamma'})}{\hbar}}
\expp^{\frac{\ci x (Q_\gamma + Q_{\gamma'})}{2 \sigma \hbar}}
\delta \left( T - \frac{T_\gamma + T_{\gamma'}}{2} \right) \right\rangle
\end{equation}
The sum is over all periodic orbits of the open system, and we have
included the term with the parametric velocities that arise from 
a variation of the first parameter $X$. In the semiclassical regime
a very small magnetic field leads to a transition from from GOE to
GUE statistics, and the influence of the magnetic field on the
classical motion can be neglected. 

One ingredient in the following calculation is the distribution of the
phases $\theta_\gamma(B)$. As for the parametric velocities it is assumed
that they have a Gaussian distribution with zero mean and a variance proportional
to the period of the orbit
\begin{equation} \label{tmean}
\left\langle \theta_\gamma \right\rangle = 
0 \; , \quad \left\langle \theta_\gamma^2 \right\rangle \sim 2 Y^2 D T \; ,
\quad T \rightarrow \infty
\end{equation}
where the averages are performed over trajectories with period around $T$,
and $D$ is a constant.

For the evaluation of (\ref{ksemi2}) we have to evaluate the averages
\begin{equation} \label{theta}
\left\langle\expp^{\frac{ \ci(\theta_\gamma-\theta_{\gamma'})}{\hbar}}\right\rangle
\end{equation}
for the different kinds of correlated orbit pairs. In the diagonal approximation
one pairs orbits that are either identical or traverse the same path in the
opposite direction. In the first case the phases are identical and the
average in (\ref{theta}) is one. In the second case the phases have opposite sign,
and with the assumption of the Gaussian distribution of the phases the average
in (\ref{theta}) becomes
\begin{equation}
\left\langle\expp^{\frac{ 2\ci \theta_\gamma}{\hbar}}\right\rangle=\expp^{-y T/T_H}
\end{equation}  
where $y=4 Y^2 D T_H/\hbar^2$. The remaining evaluation of the 
diagonal approximation follows section \ref{diagopen} and the result is
\begin{equation}
\tau\expp^{-(B+M)\tau}\left(1 + \expp^{-y \tau}\right)
\end{equation}
It interpolates between the GOE and GUE results as $y$ goes from zero to infinity.

For the correlated periodic orbit pairs that give the off-diagonal contributions
to the form factor, we have to distinguish parts of the trajectory that are followed
in the same direction and parts that are followed in the opposite direction.
This depends on the ``structure'' of the orbit pair. Whereas in all previous
calculations in this article one could combine the contributions from all structures
with the same vector $\v$, one now has to look at the structures individually.
Again one has to treat the orbit stretches and the loops separately.

Let us look first at the phase differences that arise from the encounter regions. 
In each encounter region $\alpha$ there are $l_\alpha$ orbit stretches. Each
stretch contributes a value $\pm \theta_{\text{enc}}^\alpha$ to the total phase, and
the sign depends on the direction in which the stretch is traversed. 
The total phase difference between an orbit and its partner that comes
from an encounter region is then $\pm2 n_\alpha \theta_{\text{enc}}$ where $n_\alpha$
is an integer. So for each structure one has to determine the numbers $n_\alpha$
for all encounters. These numbers can be obtained from the permutation matrices.
The contribution to the average (\ref{theta}) from an encounter region $\alpha$
is a factor
\begin{equation}
\langle\expp^{\pm 2 \ci n_\alpha \theta_{\text{enc}}^\alpha / \hbar}\rangle
=\expp^{-n_\alpha^{2} y t_{\text{enc}}^\alpha/T_H}
\end{equation}

In order to calculate the phase difference from the loops one has to specify
for each structure the number of loops $N$ that are traversed in opposite
directions by an orbit and its partner. Let $t_1, \ldots,t_N$ denote the
times along these $N$ loops. Then the contribution of the loops to the
average (\ref{theta}) is a factor $\exp(-y(t_1+\dots+t_N)/T_H)$. Of course,
the times along the loops vary for different orbit pairs which have the
same structure. The final result for the average (\ref{theta}) is
\begin{equation} \label{theta2}
\langle\expp^{\frac{ \ci(\theta_\gamma-\theta_{\gamma'})}{\hbar}}\rangle
= \expp^{-\sum_\alpha n_\alpha^{2} y t_{\text{enc}}^\alpha/T_H} \; 
\expp^{-y (t_1+\dots+t_N) / T_H}
\end{equation}

In order to evaluate the contribution of orbit pairs with the same structure
to the form factor, one has to discuss how the density of self-encounters
(\ref{defw}) was derived. It was obtained by integrating over all possible
loop lengths \cite{Mue05}
\begin{equation} \label{wint}
w_{T}(\s,\u)=\frac{T}{\Omega^{L-V} \prod_{\alpha} t_{\text{enc}}^{\alpha}}
\int_{0}^{T-t_{\text{enc}}} \ud t_{L-1}
\int_{0}^{T-t_{\text{enc}}-t_{L-1}} \ud t_{L-2} \ldots
\int_0^{T-t_{\text{enc}}-t_{L-1} \ldots - t_{2}} \ud t_{1}
\end{equation}
where $t_{\text{enc}} = \sum_{\alpha}l_{\alpha}t_{\text{enc}}^{\alpha}$ is
the sum of all the encounter times. The integral is over $L-1$ loops, the
time along the remaining $L$-th loop is determined by the condition that
the sum of all loop times plus $t_{\text{enc}}$ is $T$. We can now evaluate
the effect of the average over the phase differences on the contribution
to the form factor by including (\ref{theta2}) in the integral (\ref{wint}),
and we define
\begin{align}
\tilde{w}_{T}(\s,\u) & =\frac{T
\expp^{-\sum_\alpha n_\alpha^{2} y t_{\text{enc}}^\alpha/T_H}
}{\Omega^{L-V} \prod_{\alpha} t_{\text{enc}}^{\alpha}}
\notag \\ & \quad \times
\int_{0}^{T-t_{\text{enc}}} \ud t_{L-1}
\int_{0}^{T-t_{\text{enc}}-t_{L-1}} \ud t_{L-2} \ldots
\int_0^{T-t_{\text{enc}}-t_{L-1} \ldots - t_{2}} \ud t_{1}
\expp^{-y(t_1 + \ldots + t_N)/T_H}
\end{align}
These integrals can easily be obtained, but they depend now on $N$ as well.

Finally, the effect of the escape probability in open systems and the
variation of the first non-symmetry breaking parameter is taken into
account in the same way as in the previous section. The contribution of
a particular structure to the form factor can then be written as

\begin{equation} \label{sumbreak}
K_{\text{str}}(\tau,x,y,M) = \tau \int d^{L-V} s \; d^{L-V} u \;
\frac{\tilde{z}_T(\s,\u)}{L} \expp^{\ci \s \u / \hbar} \; .
\end{equation}
where
\begin{equation}
\tilde{z}_T(\s,\u) = \tilde{w}_T(\s,\u) \; \expp^{-BT/T_H}\expp^{-MT/T_H}
\prod_{\alpha}\expp^{-(Bl_\alpha-M)(l_\alpha-1)t_{enc}^{\alpha}/T_H}
\end{equation}
The remaining calculation can be done by a computer program. The integrals
in (\ref{sumbreak}) are treated by using (\ref{intenc}). This leads again
to the condition that after an expansion of $z_T(\s,\u)$ in all encounter
times $t_{\text{enc}}^\alpha$ only those terms survive where all encounter
times have a zero exponent. All the relevant information about the
quantities $\v$, $n_\alpha$ and $N$ for the different structures can be
found in table~2 of \cite{NBMSHH06}. We list the result in table
\ref{orbittableparaopentrans}. In order that this table is not too complicated
we have summed up the contributions to the form factor from all structures
with the same vector $\v$. We defined $\tilde{y} = y \tau$.

\begin{table}[htb]
\centering
\begin{tabular}{|c|c|}
\hline
$\v$ & $K_{\v}(\tau,x,y,M)\expp^{(B+M)\tau}$\\
\hline \hline
$(2)^{1}$&$-2\expp^{-\tilde{y}}\tau^{2} 
- (2B - M)\left(\frac{1 - \expp^{-\tilde{y}}}{\tilde{y}}\right)\tau^{3}$\\
\hline \hline
\multirow{3}{*}{$(2)^{2}$}&
$\left(2 + 6\expp^{-\tilde{y}} + \frac{2 - 2\expp^{-\tilde{y}}}{\tilde{y}} 
+ \tilde{y}\expp^{-\tilde{y}} + 
\frac{\tilde{y}^{2}\expp^{-\tilde{y}}}{6}\right)\tau^{3}$\\
&${}+(2B - M)\left(\frac{1 + \expp^{-\tilde{y}}}{2} + \frac{4 - 4\expp^{-\tilde{y}} 
+ 2\tilde{y} - 6\tilde{y}\expp^{-\tilde{y}}}{\tilde{y}^{2}} 
+ \frac{\tilde{y}\expp^{-\tilde{y}}}{6}\right)\tau^{4}$\\
&${}+(2B - M)^{2}\left(\frac{1 + \expp^{-\tilde{y}}}{24} + \frac{4\expp^{-\tilde{y}} 
- 4 + 2\tilde{y} + 2\tilde{y}\expp^{-\tilde{y}}}{\tilde{y}^{3}}\right)\tau^{5}$\\
\hline
$(3)^{1}$&$-\left(2 + 4\expp^{-\tilde{y}}+\frac{2 - 2\expp^{-\tilde{y}}}{\tilde{y}} 
+ \tilde{y}\expp^{-\tilde{y}}\right)\tau^{3} - (3B - M)\left(\frac{1 + \expp^{-\tilde{y}}}{3}
+\frac{2 - 2\expp^{-\tilde{y}}}{\tilde{y}}\right)\tau^{4}$\\
\hline\hline
\multirow{4}{*}{$(2)^{3}$}&$-\left(\frac{50\expp^{-\tilde{y}}}{3} 
+ \frac{20 - 20\expp^{-\tilde{y}} + 28\tilde{y} - 48\tilde{y}\expp^{-\tilde{y}}}{\tilde{y}^{2}} 
+ 4\tilde{y}\expp^{-\tilde{y}} + \frac{\tilde{y}^{2}\expp^{-\tilde{y}}}{3}\right)\tau^{4}$\\
&${}+(2B - M)\left(2\expp^{-\tilde{y}} + \frac{4 - 4\expp^{-\tilde{y}} - 34\tilde{y} 
+ 30\tilde{y}\expp^{-\tilde{y}} - 13\tilde{y}^{2} 
+ 45\tilde{y}^{2}\expp^{-\tilde{y}}}{\tilde{y}^{3}} 
- \frac{\tilde{y}\expp^{-\tilde{y}}}{6}\right)\tau^{5}$\\
&${}+(2B - M)^{2}\left(\frac{\expp^{-\tilde{y}}}{12} 
+ \frac{192 - 192\expp^{-\tilde{y}} - 24\tilde{y} - 168\tilde{y}\expp^{-\tilde{y}} 
- 21\tilde{y}^{2} - 51\tilde{y}^{2}\expp^{-\tilde{y}} - 4\tilde{y}^{3} 
+ 5\tilde{y}^{3}\expp^{-\tilde{y}}}{3\tilde{y}^{4}}\right)\tau^{6}$\\
&${}-(2B - M)^{3}\left(\frac{768 - 768\expp^{-\tilde{y}} 
- 384\tilde{y} - 384\tilde{y}\expp^{-\tilde{y}} + 56\tilde{y}^{2} 
- 56\tilde{y}^{2}\expp^{-\tilde{y}} 
+ 4\tilde{y}^{3} + 4\tilde{y}^{3}\expp^{-\tilde{y}} + \tilde{y}^{4} 
- \tilde{y}^{4}\expp^{-\tilde{y}}}{24\tilde{y}^{5}}\right)\tau^{7}$\\
\hline
\multirow{4}{*}{$(2)^{1}(3)^{1}$}&
$\left(16\expp^{-\tilde{y}} +
\frac{24 - 24\expp^{-\tilde{y}} + 44\tilde{y} 
- 68\tilde{y}\expp^{-\tilde{y}}}{\tilde{y}^{2}} 
+ \frac{10\tilde{y}\expp^{-\tilde{y}}}{3}\right)\tau^{4}$\\
&${}-4B\left(\frac{2\expp^{-\tilde{y}}}{3} 
+ \frac{20 - 20\expp^{-\tilde{y}} - 22\tilde{y} 
+ 2\tilde{y}\expp^{-\tilde{y}} - 11\tilde{y}^{2} 
+ 23\tilde{y}^{2}\expp^{-\tilde{y}}}{\tilde{y}^{3}}\right)\tau^{5}$\\
&${}
+M\left(\expp^{-\tilde{y}} + \frac{36 - 36\expp^{-\tilde{y}} 
- 36\tilde{y} - 17\tilde{y}^{2} 
+ 35\tilde{y}^{2}\expp^{-\tilde{y}}}{\tilde{y}^{3}}\right)\tau^{5}$\\
&${}-(2B - M)(3B - M)\left(\frac{12 - 12\expp^{-\tilde{y}} 
- 6\tilde{y} - 6\tilde{y}\expp^{-\tilde{y}} 
- \tilde{y}^{2} + \tilde{y}^{2}\expp^{-\tilde{y}}}{\tilde{y}^{3}}\right)\tau^{6}$\\
\hline
\multirow{2}{*}{$(4)^{1}$}&$-\left(2\expp^{-\tilde{y}}
+\frac{4 - 4\expp^{-\tilde{y}} + 16\tilde{y} 
- 20\tilde{y}\expp^{-\tilde{y}}}{\tilde{y}^{2}}\right)\tau^{4}$\\
&${}+(4B - M)\left(\frac{12 - 12\expp^{-\tilde{y}} 
- 6\tilde{y} - 6\tilde{y}\expp^{-\tilde{y}} 
- 9\tilde{y}^{2} + 9\tilde{y}^{2}\expp^{-\tilde{y}}}{2\tilde{y}^{3}}\right)\tau^{5}$\\
\hline
\end{tabular}
\caption{Contribution of different types of orbit pairs to the form factor $K(\tau,x,y,M)$.}
\label{orbittableparaopentrans}
\end{table}

Finally, we sum over all contributions $K_{\v}(\tau,x,y,M)$ in table \ref{orbittableparaopentrans}
and add the diagonal approximation. This gives the expansion of the form factor up to order $\tau^5$
\begin{align} \label{general}
& K(\tau,x,y,M) \notag \\
& \approx
\expp^{-(B + M)\tau} \left\{\left[1 + \expp^{-y \tau}\right]\tau 
- 2\expp^{-y \tau}\tau^{2} + \left[2\expp^{-y \tau} 
+ \frac{y^2 \tau^2\expp^{-y \tau}}{6} - \left(2B - M\right)
\left(\frac{1 - \expp^{-y \tau}}{y \tau}\right)\right]\tau^{3} \right. \notag \\
& \quad -\left[\frac{8\expp^{-y \tau}}{3} + \frac{2y \tau\expp^{-y \tau}}{3} 
+ \frac{y^2 \tau^2\expp^{-y \tau}}{3} + M\left(\frac{1 + \expp^{-y \tau}}{6}\right) 
- \left(2B - M\right)\left(\frac{y \tau\expp^{-y \tau}}{6}\right)\right. \notag \\
& \left.\quad-2B\left(\frac{4 - 4\expp^{-y \tau} - y \tau 
- 3y \tau\expp^{-y \tau}}{y^2 \tau^2}\right) + 4M\left(\frac{1 - \expp^{-y \tau} 
- y \tau\expp^{-y \tau}}{y^2 \tau^2}\right)\right]\tau^{4} \notag \\
& \quad +\left[4\expp^{-y \tau}+\frac{4y \tau\expp^{-y \tau}}{3}+\frac{y^2 \tau^2\expp^{-y \tau}}{3} 
+ \frac{y^4 \tau^4\expp^{-y \tau}}{120} + \left(4B 
- 3M\right)\frac{\expp^{-y \tau}}{3} \right. \notag \\
&\quad-\left(2B - M\right)\frac{y \tau\expp^{-y \tau}}{6} 
+ 8B\left(\frac{6\expp^{-y \tau} - 6 + y \tau + 5y \tau\expp^{-y \tau} 
+ 2y^2 \tau^2\expp^{-y \tau}}{y^3 \tau^3}\right) \notag \\
&\quad+M\left(\frac{52 - 52\expp^{-y \tau} + 2y \tau - 54y \tau\expp^{-y \tau} 
+ y^2 \tau^2 - 29y^2 \tau^2\expp^{-y \tau}}{y^3 \tau^3}\right) \notag \\
&\left.\left.\quad+\left(2B - M\right)^{2}\left(\frac{1 
+ \expp^{-y \tau}}{24} + \frac{4\expp^{-y \tau} - 4 + 2y \tau 
+ 2y \tau\expp^{-y \tau}}{y^3 \tau^3}\right)\right]\tau^{5} + \ldots \right\}
\end{align}
This result encompasses all previous results in this article, up to order $\tau^5$. In order
to compare it to the result of RMT, we have to expand the exponentials $\expp^{-y \tau}$ in
$\tau$ as discused in section~\ref{sect_rmt}. If one does this, one obtains an expansion
that agrees with all terms in (\ref{rmtall}).

\section{Conclusions}

In this article we extended previous work to include open systems.
We considered correlation functions of the Wigner time delay, which,
because it can be expressed in terms of the periodic orbits of the repeller,
can be treated in a similar way to previous calculations.
Compared to the closed system case,
the calculation must be performed by taking into account the average
probability of survival of periodic orbits.  A small, but important correction
comes from the encounter regions, where, because the traversals are
spatially close, the probability that each survives is essentially
the probability that the encounter region survives \cite{HMBH06}. 
By incorporating this correction we derive the open form factor
up to 9th order for the case without
time reversal symmetry and up to 7th order for the case with time reversal symmetry.
These results showed exact agreement with the small $\tau$ expansion of the GUE and 
GOE results respectively.  When the number of open channels $M$ is set to zero,
the system is closed, and these results agree with the terms in the spectral
form factor expansion.

We also took into account the effects of a non-symmetry breaking external
parameter.  The parametric form factor involves comparing the time
delay at two diffent values of this parameter, correctly rescaled.  The
semiclassical calculation assumes a Gaussian
distribution of the parametric velocities of long orbits and 
again requires a small correction from the encounter regions.
This correction comes from the fact that the traversals of the
encounter region are correlated, affecting the Gaussian averaging.
We derived the open parametric form factor again up to 9th and 7th order
for the GUE and GOE cases and found agreement with the RMT expansions.
When the scaled parameter difference $x$ was set to zero, effectively 
excluding parametric correlations, we recover the results of the form factor
for the time delay without an external parameter.
Likewise, closing the system by setting the number of open channels $M$ to zero
allows us to recover previous parametric results \cite{NBMSHH06,KS06}.

The final, and most general, case we considered also allowed for a second
symmetry breaking parameter $y$, describing a magnetic field.  This parameter
gives an additional term that depends on the time an orbit and
its partner spend travelling in opposite directions.  Semiclassically, the calculation
is complicated by the need to consider each ``structure'' separately, as
it now depends on the number of loops and encounter traversals that are
followed in opposing directions \cite{NS06,NBMSHH06}.  We obtain a semiclassical
result up to 5th order in $\tau$, from which we can recover the parametric form factor
for the time delay
by taking the limit $y=0$ to get the GOE case and $y\to\infty$ to get the GUE case.
For fixed $y$ we recover the small $\tau$ expansion of the RMT result.
Also by removing the parametric correlation $x=0$, and closing the system $M=0$, we
recover the GUE-GOE transition result \cite{NS06,NBMSHH06}.

The semiclassical calculation for all these cases agrees with the small $\tau$
expansion of the appropriate RMT result. The limitations of the calculation
are similar to that for the spectral form factor. A main difficulty is to show that
there are no other semiclassical contributions that survive the semiclassical limit
and would lead to a deviation from the RMT results. We also assume that the average
over survival times and Gaussian distributed parametric velocities can be done
indepentently of the sum over orbits.  Another open point
concerns the region $\tau>1$, in which the random matrix expressions for
$K(\tau,x,y,M)$ have a different functional form. 

\section*{Acknowledgements}

The authors would like to thank EPSRC for financial support.

\end{document}